\documentclass[12pt]{article}
\usepackage{cite,epsfig,amssymb,amsmath,amstext,graphicx,color,subfigure,xcolor,mathrsfs,tikz}
\evensidemargin \oddsidemargin

\usetikzlibrary{decorations.pathmorphing}
\usepackage{latexsym}
\usepackage{hyperref}
\usepackage{bbm}
\hypersetup{colorlinks,%
  citecolor=blue,%
  linkcolor=black,%
  pdftex}
\usepackage[numbers,sort&compress,square]{natbib}
\usepackage{varioref}
\usepackage{slashed}
\usepackage[normalem]{ulem}
\usepackage[makeroom]{cancel}
\usepackage{pdfsync}
\usepackage{kotex}
\usepackage{soul}
\usepackage{pgfplots}

\begin{document}

~
\vspace{5mm}

\begin{center}

{{\Large \bf Asymptotic Weyl Symmetry and Its Anomaly in a Curved Spacetime }}
\\[5mm]

Jeongwon Ho$^1$, ~~O-Kab Kwon$^1$, ~~Sang-A Park$^1$, ~~Sang-Heon Yi$^2$  \\[2mm]
{\it $^1$Department of Physics,~Institute of Basic Science, Sungkyunkwan University, Suwon 16419, Korea}  
\\
{\it $^2$Physics Department  Natural Science Research Institute
University of Seoul, Seoul 02504 Korea } 
\\[2mm]
{\it freejwho@gmail.com, ~okab@skku.edu, ~psang314@gmail.com, ~shyi704@gmail.com}
\end{center}
\vspace{15mm}

\begin{abstract}
\noindent
  We explore an unusual symmetry  in a field theory on a specific (1+1)-dimensional curved spacetime, which has an interesting interpretation as an approximate asymptotic Weyl symmetry. Unlike the conventional Weyl symmetry, the boundary term under the variation plays a crucial role in understanding for its anomaly. After converting a two-dimensional field theory on curved spacetime  to an inhomogeneous field theory, we  obtain the vacuum expectation value of the energy-momentum tensor. Then, we show  the existence of an Unruh-like effect  in the bubble wall expansion at the zero temperature.

\end{abstract}


\thispagestyle{empty}
 
\newpage
 \tableofcontents

\section{Introduction }
\label{sec1}

A classical field theory whose action is invariant under local Weyl transformations exhibits a vanishing trace of the energy--momentum tensor, $T^\mu_{\ \mu}=0$. More precisely, under a Weyl transformation $g_{\mu\nu}(\pmb{x})\to e^{2\sigma(\pmb{x})}g_{\mu\nu}(\pmb{x})$, given by
\begin{equation} \label{actionWeylTr}
\delta_\sigma S = \int d^dx\,\sqrt{-g}\,\sigma(\pmb{x})\,T^\mu_{\ \mu}, \qquad \pmb{x}\equiv(t,\vec{x})\,,
\end{equation}
Weyl invariance at the classical level requires $T^\mu_{\ \mu}=0$. At the quantum level, this symmetry is generically broken, leading to the well-known trace anomaly, $\langle T^\mu_{~ \mu}\rangle \neq 0$ \cite{Duff:1993wm}.

On the other hand, 
the study of symmetries restricted to the asymptotic boundary of spacetime has provided  profound insights. 
In four-dimensional general relativity, the asymptotic symmetry group on the asymptotically flat spacetime is not the Poincar\'e group but the infinite-dimensional BMS group \cite{Bondi:1962px, Sachs:1962wk, Strominger:2017zoo}. 
In the context of 2D conformal field theory, the infinite-dimensional Virasoro symmetry is realized as the asymptotic symmetry group of the 3D AdS bulk~\cite{Brown:1986nw}. 
Especially, the asymptotic symmetry analysis in bulk AdS$_{3}$  is regarded as the precursor of the modern AdS/CFT correspondence.
In these canonical examples, the asymptotic symmetry is understood as an exact remnant or subgroup of classical bulk symmetry. The quantum anomaly is then interpreted as a modification of this symmetry algebra, such as a central extension.

This paper, however, addresses a bit different and subtle scenario: We investigate the quantum effect of a (1+1)-dimensional scalar field theory, which enjoys an asymptotic symmetry but is  {\it not}  Weyl-invariant with non-minimal coupling to gravity, described by the action 
\begin{align}\label{actionFTCS}
	S_{\mathrm{FTCS}} &= \int_{\mathcal{M}} d^2x\,\sqrt{-g}\,\bigg(  -\frac{1}{2}g^{\mu\nu}\nabla_\mu\phi\nabla_\nu\phi  -\frac{1}{2}m_0^2\phi^2  -\frac{1}{2}\xi\mathcal{R}\phi^2\bigg)\,,
\end{align}
where $m_0$ and $\xi$ are constants, and $\mathcal{R}$ denotes the curvature scalar. 
In our previous study~\cite{Ho:2024kzr}, we showed that the asymptotic behavior 
of \(\langle T^{\mu}{}_{\mu} \rangle\) coincides with the well--known universal 
trace--anomaly form of the conformal scalar field theory.
 This interesting coincidence suggests the possible existence of a hidden symmetry that has so far remained unexplained. While several works have investigated the generalization of the trace anomaly to non-conformal theories (see, for example, \cite{Ferrero:2023unz}), the present study offers an alternative viewpoint toward understanding this phenomenon, based on the presence of an asymptotic Weyl symmetry (AWS), which will be explained in later sections.

We established  in \cite{Kwon:2022fhv,Ho:2022omx} that the field theory on curved spacetime (FTCS) \eqref{actionFTCS} on a conformally flat metric $ds^2 = e^{2\omega(x)}\eta_{\mu\nu}dx^\mu dx^\nu$ is classically converted to an inhomogeneous field theory (IFT) on (1+1)-dimensional flat spacetime:
\begin{align}\label{actionIFT}
	S_{\rm IFT} = \int d^2 x \bigg( -\frac{1}{2} \eta^{\mu\nu} \partial_\mu \phi \partial_\nu \phi - \frac{1}{2} m_{\rm eff}^2(x) \phi^2 \bigg)\,.
\end{align}
In this framework, geometric information is encoded into the position-dependent effective mass
\begin{align}\label{effMassSq}
	m_{\rm eff}^2(x) = e^{2 \omega(x)}\big( m_0^2 + \xi \mathcal{R} \big) .
\end{align}
We suggested to extend this conversion  to the quantum level, allowing for a consistent quantization, e.g., using Hadamard regularization techniques from the FTCS framework to compute correlation functions  in the IFT \cite{Ho:2022omx}.

Since the IFT action~\eqref{actionIFT} is defined on a fixed flat background, the AWS cannot act on the metric $\eta_{\mu\nu}$. Moreover, as the scalar field $\phi$ remains invariant, the symmetry must instead be realized as an internal symmetry acting on the inhomogeneous parameter of the IFT—the effective mass $m_{\rm eff}^2(x)$. In this sense, the AWS is regarded as a hidden symmetry from the IFT perspective.

This paper is organized as follows.
In Section \ref{sec2}, we review the conversion between FTCS and IFT by extending our previous work to a non-supersymmetric background. 
Section \ref{sec3} introduces a new asymptotic symmetry, which we refer to as the AWS, and examines its anomalous features. 
In Section \ref{sec4}, we compute the renormalized Hadamard two-point function and the vacuum expectation value (VEV) of the energy--momentum tensor, and we discuss the physical implications of these results. 
Finally, Section \ref{sec5} presents concluding remarks and possible future directions.

\section{FTCS–IFT Conversion Beyond SUSY Backgrounds}\label{sec2}

We have investigated the classical conversion between  FTCS \eqref{actionFTCS} 
and  IFT~\eqref{actionIFT} in $(1+1)$ dimensions, and applied this relation to translate 
the quantum effects of the FTCS into those of the IFT~\cite{Kwon:2022fhv, Ho:2022omx, Ho:2024kzr}. 
In these works, our analysis mainly relied on a supersymmetric background geometry. 
In the present paper, we instead employ a non-supersymmetric geometry that contains a single 
parameter interpolating the previously used supersymmetric geometry and the Minkowski 
spacetime. Our goal is to focus on the AWS and its associated anomaly, which were overlooked 
in our previous studies, and to interpret those quantum effects in IFT.

\subsection{Review on conversion between FTCS and IFT}

Since the conversion between FTCS and IFT constitutes a key element of our framework, we begin by briefly revisiting its conceptual basis.
A common objection to our approach is that two actions with different underlying symmetry groups cannot describe the same physical system. While this argument is reasonable in a generic context, it does not strictly apply to the present case.
Our starting point in converting the FTCS (in a specific conformal gauge) to the IFT is the same classical equation of motion and the same initial conditions for the scalar field $\phi$. Consequently, the classical solutions of $\phi$ are identical in both formulations, even though the actions differ. The discrepancy in their symmetry structures should thus be viewed not as a physical inconsistency, but rather as reflecting distinct interpretations—or domains of applicability—of the same underlying solutions.

Our central claim in previous studies is that, although no fully consistent quantization scheme for the IFT is yet known, the quantization of the IFT can be partially understood through its conversion from the FTCS.\footnote{We do not claim that the full quantum IFT is equivalent to the full quantum FTCS, even for the scalar field, since the conversion is performed within a specific conformal gauge. While the FTCS may appear more restricted due to diffeomorphism constraints, the conversion nevertheless captures certain quantum aspects of the IFT. In what follows, we focus on the intersection regime in which quantum features of the IFT can be meaningfully interpreted through this correspondence.}
This viewpoint is motivated by the field quantization formalism on the solution space in curved spacetime~\cite{Wald:1995yp}. Based on the same solution space, we argued that the quantization of the IFT can, at least in part, be inferred from its correspondence with the FTCS.

In the specific scalar field example, the effective mass $m_{\rm eff}^2$ in the IFT can be decomposed into two contributions $m_0^2$ and $\xi \mathcal{R}$ in \eqref{effMassSq} from the FTCS perspective. 
From the viewpoint of the IFT, this division may appear artificial; nevertheless, it serves as a practical prescription for the IFT quantization scheme, rather than as a unique or unambiguous construction.

In particular, one of our proposals in this framework is that the energy–momentum tensor in the IFT can be obtained from that in the FTCS by adopting the same algebraic vacuum state in both formulations, thereby defining a common Fock space. We continue to employ this prescription in the present work and interpret our results accordingly.
Although different decompositions of $m_{\rm eff}^2$ may correspond to distinct quantization schemes, it is plausible that certain physical consequences remain universal and independent of such choices. More concretely, while the algebraic vacuum\footnote{These states possess considerable freedom and must ultimately be restricted by physical criteria.} in the IFT and FTCS can be chosen to coincide by construction, this does not necessarily guarantee its physical appropriateness in the IFT framework. However, in the absence of any clear symmetry principle, constraint, or physical argument forbidding this choice, it is reasonable to adopt this identification and explore its physical implications. A complete resolution of this issue lies beyond the scope of the present work and will require further investigation.

In the following, we introduce a one-parameter family of background metrics, specified by 
$\alpha$, which continuously interpolates between the Minkowski spacetime ($\alpha = 1$) 
and the supersymmetric background studied in our previous works~\cite{Ho:2022omx,Ho:2024kzr} 
at $\alpha = 0$. The non-supersymmetric cases ($0<\alpha<1$) are free from the curvature singularity that appears in the left asymptotic limit of $x \rightarrow -\infty$     in the  supersymmetric background. Upon converting the FTCS to the IFT description, this geometric modification manifests as a distinctive feature of the scalar IFT--namely, the emergence of two different asymptotic mass parameters, $m_{\ell}^2$ and $m_{r}^2$   of a single scalar IFT.  Here, $m_{\ell}$ denotes the mass in the left asymptotic limit, while $m_{r}^{2}$ 
represents the mass in the right asymptotic limit. The supersymmetric background in the FTCS corresponds to the special case where $m_{\ell}^2 = 0$ in the IFT framework.

\subsection{A static background}

In \cite{Ho:2022omx,Ho:2024kzr}, we studied the FTCS model \eqref{actionFTCS} describing a free massive scalar field propagating on a two-dimensional curved background   described by   the conformally flat metric depending only on $x$:
\begin{align}\label{metric}
	ds^{2} &= e^{2\omega(x)}\big(-dt^{2}+dx^{2}\big)\,.
\end{align}
In this case, the curvature scalar can be written as
\begin{align}\label{curvatureScalar}
	\mathcal{R} &=-2e^{-2\omega(x)}\omega''(x)\,.
\end{align}
Varying the action \eqref{actionFTCS} with respect to the scalar field and the metric yields the field equation
\begin{align}\label{eomFTCS}
\big( -\Box + m_0^2 + \xi \mathcal{R} \big) \phi = 0, \qquad
\Box = \frac{1}{\sqrt{-g}} \partial_\mu \big( \sqrt{-g} \, g^{\mu\nu} \partial_\nu \big),
\end{align}
   where $\xi>0$ and $m_{0}^{2}>0$,
and the classical energy--momentum tensor
\begin{align} \label{classicalStress}
T_{\mu\nu}= \nabla_{\mu}\phi\,\nabla_{\nu}\phi -\tfrac{1}{2}g_{\mu\nu}\Big[(\nabla\phi)^2 + m_{0}^2\, \phi^2\Big]  +  \xi  \Big(- \nabla_{\mu}\nabla_{\nu} + g_{\mu\nu}\nabla^{2} \Big)\phi^{2},
\end{align}
respectively. In the specific conformal gauge~\eqref{metric},  the equations of motion of the scalar field becomes 
\begin{equation} \label{waveEq}
	\big(-\eta^{\mu\nu}\partial_{\mu}\partial_{\nu} + m^{2}_{\rm eff}(x)\big) \phi =0\,,  
\end{equation}
where $m_{\textrm{eff}}^{2}(x)$ has same expression in \eqref{effMassSq}.
The same equation of motion can also be read from the IFT action \eqref{actionIFT}.

\subsection{Non-supersymmetric spacetime background}

\begin{figure}[ht]
  \centering
\includegraphics{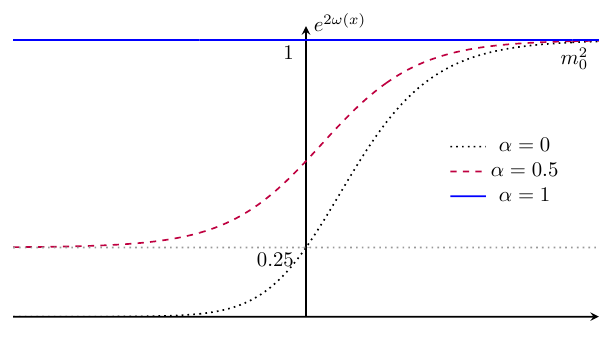}
  \caption{Plot of $e^{2\omega(x)}$ for $b=1$ and $\alpha=0,0.5,1$.}
  \label{fig:metric}
\end{figure}

Our static metric, which is a one-parameter extension of the supersymmetric metric with choice of $b>0$, is taken by 
\begin{align}\label{metricModified}
	e^{\omega(x)} &=\frac{1+\alpha e^{-bx}}{1+e^{-bx}}\,,\qquad 0\leq\alpha \leq1 \,.
\end{align}
The FTCS action~\eqref{actionFTCS} cannot be supersymmetrized on the 
background metric~\eqref{metricModified} for \(0 <\alpha < 1\). For the details see Appendix~\ref{appendixB}.
 It would be plausible to assume that the physical quantities would be a smooth or continuous with respect to  the parameter $\alpha$ except the limit\footnote{Though the limit $\alpha \to 0$ may be  continuous or smooth in the IFT context, we allow its non-continuity as a possibility.} $\alpha \to 0$.
In Fig.~\ref{fig:metric}, we plot the conformal factor $e^{2\omega(x)}$ for three representative cases: the supersymmetric background $(\alpha = 0)$, an intermediate geometry $(\alpha = 0.5)$, and the flat spacetime limit $(\alpha = 1)$.

We now summarize the global and qualitative features of $m_{\rm eff}^{2}(x)$ arising from the background metric \eqref{metricModified}.

\paragraph{Global behavior:}
The dimensionless effective mass square profile,
\begin{align}\label{effMassSq1}
\frac{m_{\textrm{eff}}^{2}(x)}{m_{0}^{2}} &=\bigg(\frac{1+\alpha e^{-bx}}{1+e^{-bx}}\bigg)^{2} +\frac{2\xi b^{2}}{m_{0}^{2}}\bigg(\frac{e^{-bx}}{(1+e^{-bx})^{2}}-\frac{\alpha e^{-bx}}{(1+\alpha e^{-bx})^{2}}\bigg)\,,
\end{align}
interpolates smoothly between two asymptotic limits,
\begin{align}\label{effMassSqAsympt}	\frac{m_{\textrm{eff}}^{2}(x)}{m_{0}^{2}}\to \bigg\{
	\begin{array}{ll}
	1 & x\to+\infty\\
	\alpha^{2} & x\to-\infty
	\end{array}\,.
\end{align}
This position-dependent mass profile exhibits a single smooth transition between the two  asymptotic    constant mass values, $m_{\ell}^2 = m_0\alpha^2$ in the left and $m_r^2 = m_0^2$ in the right.  From \eqref{effMassSq1}, one can also see that global bounds of $m_{\textrm{eff}}^{2}$ exist for all $x$. 
In particular, a sufficient condition for strict positivity of the inhomogeneous mass parameter, $m^{2}_{{\rm eff}}(x)$,   is
\begin{align}\label{suffcon}	\alpha^{2}>\frac{\xi b^{2}}{2m_{0}^{2}}\,.
\end{align}
This inequality guarantees that the profile remains positive definite throughout the entire spatial domain. Though $m^{2}_{\rm eff}(x) < 0$ for some range of $x$ may be an interesting possibility of IFT, we will focus on $\alpha^{2}>\frac{\xi b^{2}}{2m_{0}^{2}}$ in the whole range of $x$ in the following.

To figure out the properties of $m_{{\rm eff}}^{2}(x)$,   we  differentitate \eqref{effMassSq1} with respect to $x$, leading to
\begin{align}\label{effMassSqDiff}
	\frac{d}{dx}\bigg(\frac{m_{\textrm{eff}}^{2}}{m_{0}^{2}}\bigg) &=2be^{-bx}\bigg[\frac{(1-\alpha)(1+\alpha e^{-bx})}{(1+e^{-bx})^{3}} -\frac{\xi b^{2}}{m_{0}^{2}}\bigg(\frac{1-e^{-bx}}{(1+e^{-bx})^{3}}-\frac{\alpha(1-\alpha e^{-bx})}{(1+\alpha e^{-bx})^{3}}\bigg)\bigg]\,.
\end{align}
For $0<\alpha<1$, the first term in the bracket is always positive, whereas the second term, proportional to $\xi$, can change its sign within the intermediate region.  For small $\xi b^{2}/m_{0}^{2}$ the profile increases monotonically from $\alpha^{2}$ to $1$. As $\xi b^{2}/m_{0}^{2}$ grows, a shallow minimum and later a weak maximum emerge, reflecting local compensation between the two contributions in \eqref{effMassSq1}.

\begin{figure}[ht]
  \centering
\includegraphics{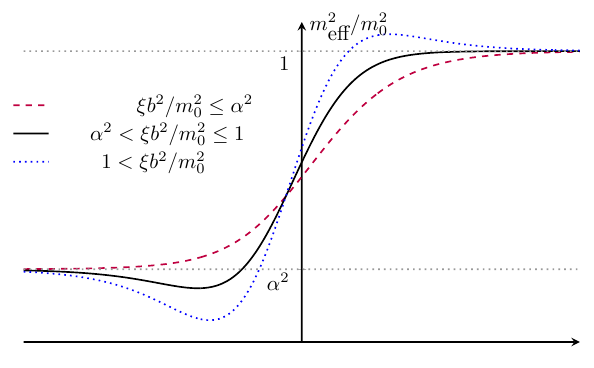}
  \caption{Plot of $m_{\textrm{eff}}^{2}/m_{0}^{2}$ for $\alpha=0.5$ and $\xi b^{2}/m_{0}^{2}=0.1,1,1.9$.}
  \label{fig:effectiveMassSq}
\end{figure}

\paragraph{Qualitative behavior:}
For $\xi > 0$ and $m_{0}^{2} > 0$, the number of extrema is controlled by the curvature-induced contribution proportional to $2\xi b^{2}/m_{0}^{2}$ in \eqref{effMassSq1}:
\begin{align}
\begin{cases}
\displaystyle \frac{\xi b^{2}}{m_{0}^{2}} \le \alpha^{2}: &
\text{monotonic increase (no extrema)},\\
\displaystyle \alpha^{2} < \frac{\xi b^{2}}{m_{0}^{2}} \le 1: &
\text{one minimum in the intermediate region},\\
\displaystyle \frac{\xi b^{2}}{m_{0}^{2}} > 1: &
\text{one minimum and one maximum.}
\end{cases}
\end{align}
Hence, the coupling $\xi b^{2}/m_{0}^{2}$ controls whether additional extrema emerge beyond the smooth interpolation between the two asymptotic masses. These features are illustrated in Fig. \ref{fig:effectiveMassSq}.

\begin{figure}[ht]
  \centering
\includegraphics{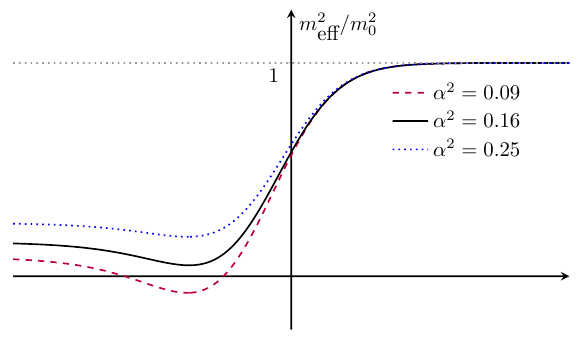}
  \caption{Plot of $m_{\textrm{eff}}^{2}/m_{0}^{2}$ for $\alpha=0.5$ and $\xi b^{2}/m_{0}^{2}=0.1,1,1.9$.}
  \label{fig3}
\end{figure}

For the following analysis, we set
\begin{align}\label{paracon}
	\frac{\xi b^{2}}{m_{0}^{2}}=1\,,
\end{align}
which provides asymptotic behavior of $m_{\mathrm{eff}}^{2}(x)$ allowing the AWS. 
For the overall positivity of $m_{\mathrm{eff}}^{2}(x)$, as we see in \eqref{suffcon}, 
we focus on a representative case given by 
$\frac{1}{2}<\alpha^{2}<1$. 
 See Fig. \ref{fig3}.

\section{Asymptotic Weyl Symmetry}
\label{sec3}

The FTCS action \eqref{actionFTCS} does not enjoy conventional Weyl symmetry, as is
already clear from the non–vanishing trace of the classical energy–momentum tensor,
\begin{equation}
  T^{\mu}{}_{\mu} = \bigl(\xi \Box - m_0^2\bigr)\phi^2 \,,
\label{ClassTrEM}
\end{equation}
obtained from \eqref{classicalStress}. Nevertheless, we shall show that, for a suitable choice of
parameters, the theory admits an \emph{asymptotic} version of Weyl symmetry in the right
asymptotic region $x\to +\infty$.

As already mentioned in Section \ref{sec2}, we focus on an asymptotically flat background in which
the effective mass $m_{\rm eff}^2(x)$ develops a nearly flat profile in the right asymptotic
region $x\to +\infty$ (see Fig.~\ref{fig:effectiveMassSq}). In the conformally flat metric
\eqref{metric}, this ``nearly flat'' behaviour is encoded in the asymptotic expansion
\begin{equation}
\label{asymptoticmetric}
  e^{2\omega(x)} = 1 + g_1 e^{-bx} + \mathcal{O}(e^{-2bx}) \,.
\end{equation}
In our static metric in~\eqref{metricModified}, the coefficient $g_{1}$ is given by $g_{1} = 2(\alpha-1)$. 

\subsection{Classical AWS}

Let us consider a Weyl rescaling of the form
\begin{align}
\label{asymptoticWeyl1}
  g_{\mu\nu} \longrightarrow \tilde{g}_{\mu\nu}
   = e^{2\sigma(t,x)} g_{\mu\nu}
   = g_{\mu\nu} + \delta g_{\mu\nu}\,,
  \qquad
  e^{2\sigma(t,x)} = 1 + a_1 e^{-bx} + \mathcal{O}(e^{-2bx}) \,,
\end{align}
where $a_1$ is taken  as a constant for simplicity. This induces the variations
\begin{align}
  \delta \sqrt{-g} &= a_1 e^{-bx} + \mathcal{O}(e^{-2bx}) \,, \nonumber\\
  \delta\bigl(\sqrt{-g}\,\mathcal{R}\bigr)
   &= - b^2a_1\,e^{-bx} + \mathcal{O}(e^{-2bx}) \,, \nonumber
\end{align}
and hence the variation of the FTCS action takes the form
\begin{equation}
  \delta S_{\rm FTCS}
   = -\frac{1}{2} \int_{\cal M} d^2x\,
      \biggl[
        a_1 e^{-bx} \big( m_0^2 
        - \xi b^2 \big)\phi^2
        + \mathcal{O}(e^{-2bx})
      \biggr] .
\label{deltaSFTCS}
\end{equation}
If the condition \eqref{paracon} is satisfied,  the action
is invariant up to the order $e^{-bx}$.

The above discussion can be recast in terms of the energy–momentum tensor and an
associated boundary term. Varying the action with respect to the background metric yields
\begin{equation}
\label{Svar}
  \delta_g S_{\rm FTCS}
   = -\frac{1}{2}\int_{\cal M} d^2x \sqrt{-g}\,
      \Bigl[T_{\mu\nu}\,\delta g^{\mu\nu} + 2\xi\, B \Bigr] \,,
\end{equation}
where the boundary term $B$ is given by
\begin{equation}
  B \equiv \nabla_\alpha\Bigl(
         \phi^2 \nabla^\alpha(g_{\mu\nu}\delta g^{\mu\nu})
        - g_{\mu\nu}\delta g^{\mu\nu}\nabla^\alpha\phi^2
      \Bigr)
     - \nabla_\mu\Bigl(
         \phi^2 \nabla_\nu\delta g^{\mu\nu}
        - \delta g^{\mu\nu}\nabla_\nu\phi^2
      \Bigr) .
\end{equation}
For the AWS transformation \eqref{asymptoticWeyl1}, with
$\delta g_{\mu\nu} = a_1 e^{-bx} g_{\mu\nu} + \mathcal{O}(e^{-2bx})$, this reduces to
\begin{equation}\label{bdB}
  B = \bigl(\Box\phi^2 - b^2\phi^2\bigr)\,a_1 e^{-bx}
      + \mathcal{O}(e^{-2bx}) \,,
\end{equation}
and then, the variation of the action \eqref{Svar} becomes 
\begin{align}
  \delta_{\rm AWS} S_{\rm FTCS}
   = -\frac{1}{2}\int_{\cal M} d^2x \sqrt{-g}\,
       \Bigl[-a_1 e^{-bx} \Big ( T^{\mu}{}_{\mu} - \xi (\Box \phi^2 - b^2 \phi^2) \Big)
             + \mathcal{O}(e^{-2bx})
       \Bigr].
\label{deltaAWS_offshell}
\end{align}
Using \eqref{ClassTrEM}, we can see that this expression is nothing but \eqref{deltaSFTCS}. 

In a  Weyl–invariant theory the boundary term does not contribute and
$T^{\mu}{}_{\mu}$ vanishes at the  off-shell level. In our non–conformal case, by contrast, the boundary
term is essential: the classical trace~\eqref{ClassTrEM} does not vanish even at the on-shell level.  However, using \eqref{paracon}, we can check that the variation of the action given in   \eqref{deltaSFTCS} or equivalently  \eqref{deltaAWS_offshell} vanishes up to the order $e^{- bx}$.

\subsection{AWS and anomaly}

A natural question is whether the AWS holds at the quantum level. In this
subsection, we argue that the renormalization procedure breaks AWS, inducing a trace–anomaly–like expression to the VEV of the energy–momentum
tensor up to  $\mathcal{O}(e^{-bx})$.

We perform the analysis in the Euclidean path–integral formalism. The partition
function and the generating functional are defined by
\begin{equation}
  Z[g] \equiv \int \mathcal{D}\phi\, e^{-S_{\rm FTCS}[\phi;g]} \,,
  \qquad
  W[g_E] \equiv -\ln Z[g] \,.
\end{equation}
Using \eqref{Svar}, we have a generic variation of the generating functional with respect to the
metric as
\begin{equation}
\label{Wvar}
  \delta_g W
   = -\frac{1}{2}\int d^2x\,\sqrt{g}\,
     \Bigl\langle T_{\mu\nu}\,\delta g^{\mu\nu}
                 + \xi B
     \Bigr\rangle_{\rm ren} .
\end{equation}
If the path–integral measure were invariant under AWS, the right–hand side of
\eqref{Wvar} would vanish up to order $e^{-bx}$ when $\delta g_{\mu\nu}$ is chosen as in
\eqref{asymptoticWeyl1}. However, it is well known that the measure is not invariant under a
generic Weyl rescaling, and in two dimensions the corresponding anomaly takes the form
\begin{equation}
  \delta_\sigma W
   = \frac{1}{24\pi}\int d^2x\,\sqrt{g}\, \mathcal{R}\,\sigma \,,
  \qquad
  \delta_\sigma g_{\mu\nu} = 2\sigma\, g_{\mu\nu} \,.
\end{equation}
Applying this to the AWS transformation with
$2\sigma = a_1 e^{-bx} + \mathcal{O}(e^{-2bx})$, we obtain
\begin{equation}
  \Big\langle
     T^{\mu}{}_{\mu}
     - \xi\bigl(\Box\phi^2 - b^2\phi^2\bigr)
  \Big\rangle_{\rm ren}
   = \frac{1}{24\pi}\,\mathcal{R}
     + \mathcal{O}(e^{-2bx}) \,,
\end{equation}
which can be rewritten as the renormalized trace of the energy–momentum tensor,
\begin{equation}
\label{AWSAnomaly}
  \big\langle T^{\mu}{}_{\mu} \big\rangle_{\rm ren}
   = \Big\langle
       \xi\bigl(\Box\phi^2 - b^2\phi^2\bigr)
     \Big\rangle_{\rm ren}
     + \frac{1}{24\pi}\,\mathcal{R}
     + \mathcal{O}(e^{-2bx}) \,.
\end{equation}

For the tuned parameter choice \eqref{paracon} the first term on the right–hand side
of \eqref{AWSAnomaly} corresponds to  the classical trace \eqref{ClassTrEM}. In the vacuum
state that we shall specify in Section \ref{sec4} this contribution is parametrically suppressed in the
right asymptotic region, so that the renormalized trace is effectively governed by the
curvature term alone. Thus, although the model is not conformal, the asymptotic Weyl
symmetry ensures that the leading quantum trace takes the same universal form as in the
conformal case. We refer to this effect as the \emph{asymptotic Weyl anomaly}: the quantum
breaking of an approximate symmetry that holds only asymptotically.
In Section \ref{sec4}, we shall verify this picture explicitly by computing the renormalized
energy-momentum tensor via the Hadamard renormalization method for specifically chosen vacuum.

\subsection{Interpretation in the IFT description}

Finally, let us briefly interpret the AWS condition in the IFT framework introduced in
Section \ref{sec1} and Section \ref{sec2}. There we showed that the FTCS model on the conformally flat metric
\eqref{metric} is classically converted  to an inhomogeneous scalar field theory on flat
spacetime, with action \eqref{actionIFT} and position–dependent mass
\eqref{effMassSq}. In terms of the IFT variables, the effective mass admits the asymptotic
expansion
\begin{equation}
  m_{\rm eff}^2(x)
    = m_0^2
      + g_1\bigl(m_0^2 - \xi b^2\bigr)e^{-bx}
      + \mathcal{O}(e^{-2bx}) \,.
\end{equation}
Hence the $\mathcal{O}(e^{-bx})$ contribution to $m_{\rm eff}^2(x)$ vanishes precisely when the
AWS condition \eqref{paracon} is satisfied. From the IFT perspective, this means that
the  parameter tuning that allows AWS in the FTCS description
corresponds to a highly flattened mass profile in the right asymptotic region. Without our proposal for 
FTCS–IFT conversion, this symmetry aspect of the tuning would remain obscure in the
IFT description. 
We would like to interpret AWS as a hidden in IFT. 
Of course, this does not imply that a hidden symmetry necessarily exists in every IFT.

\section{Quantum Energy-momentum Tensor}
\label{sec4}

As emphasized in Section~\ref{sec2}, the background metric is asymptotically flat on both sides for $0 < \alpha \le 1$. In this section, we focus on the right asymptotic region $x \to +\infty$ within the parameter regime \eqref{paracon}, where the AWS discussed in Section~\ref{sec3} is realized. Our aim is to compute the renormalized energy--momentum tensor in a specified vacuum state and to explicitly confirm the asymptotic Weyl anomaly previously derived in a formal manner in Section~\ref{sec3}.

\subsection{Renormalized Hadamard two-point function}
\label{subsec4.1}

We begin by analyzing the FTCS action~\eqref{actionFTCS} on the background 
geometry~\eqref{metricModified}, focusing on the regime in which AWS is 
present in the right asymptotic limit. To investigate quantum effects in this
region, we compute the VEV of the 
energy--momentum tensor. Following the approach developed in our previous 
work~\cite{Ho:2024kzr}, we employ the Hadamard regularization method to 
construct the renormalized two-point function.

\begin{figure}[ht]
  \centering
\includegraphics{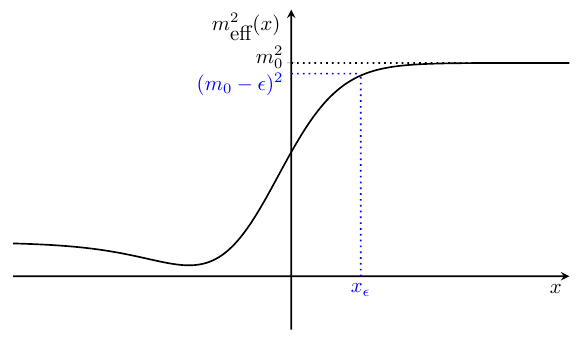}
  \caption{Plot of the effective mass profile $m_{\mathrm{eff}}^{2}(x)$ for $\alpha = 0.5$ and 
$\xi b^{2}/m_{0}^{2} = 1$, satisfying the condition~\eqref{paracon}. From the expression in~\eqref{effMassSq1}, one finds that 
$\epsilon \sim \mathcal{O}(e^{-2 b x_\epsilon})$.  
Hence, the contribution of $c_{\omega}$ is subleading and can be safely neglected in our analysis.}
  \label{fig4}
\end{figure}

As a first step, we evaluate the VEV of the product of 
two field operators at a point $x_{\epsilon}$, which is located slightly 
away from the right asymptotic region. For concreteness, we approximate
\begin{align}\label{twopf}
\omega_{\mathrm{R}}^{\epsilon}\!\left(\phi(\pmb{x})\,\phi(\pmb{x}')\right)
\simeq {}_{\mathrm{R}}^{\epsilon}\!\langle 0 \,|\, 
\phi(\pmb{x})\,\phi(\pmb{x}') \,|\, 0 \rangle_{\mathrm{R}}^{\epsilon},
\end{align}
where $\omega_{\mathrm{R}}^{\epsilon}$ denotes the Hadamard state defined 
at $x_{\epsilon}$. The state $|0\rangle_{\mathrm{R}}^{\epsilon}$ represents 
an approximate vacuum, specified by the annihilation conditions associated 
with the operators $b_{k}$ and $c_{\omega}$ defined at $x_{\epsilon}$. 
For further details, see (2.52), (2.53) and Appendix B of \cite{Ho:2024kzr}. Fig. \ref{fig4} provides an additional plot illustrating the situation under consideration.
In practice, if one keeps \eqref{twopf} only up to terms of order 
$e^{-b x}$, the contribution from the $c_{\omega}$ sector
can be neglected. In what follows we therefore evaluate the right-hand 
side of \eqref{twopf} using only the $b_{k}$ operators\footnote{For more rigorous canonical approach, we can adopt the Weyl-Titchmarsh-Kodaira theory~\cite{Kubota:2024wgx, Kubota:2025avm}. }.

In order to apply the Hadamard regularization method to the VEV of the
two-point function discussed above, we first compute the (positive-frequency)
Wightman function $G^{+}(\pmb{x},\pmb{x}')$ defined by
\begin{align}
(\Box - m_0^{\,2} - \xi\,\mathcal{R})\,
G^{+}(\pmb{x},\pmb{x}')
= -\frac{1}{\sqrt{-g}}\,
\delta(t-t')\,\delta(x-x').
\end{align}
The Wightman function can be expressed in terms of mode functions as
\begin{equation}\label{G+xx}
G^{+}(\pmb{x},\pmb{x}')
= \int d\mu_{\eta}\,
u_{\eta}(\pmb{x})\,u_{\eta}^{*}(\pmb{x}') ,
\end{equation}
where $d\mu_{\eta}$ denotes the measure associated with the mode index~$\eta$,
and $u_\eta(\pmb{x})$ satisfies the field equation \eqref{eomFTCS} or
\eqref{waveEq}. Exploiting time-translation invariance, we write
\begin{align}\label{uw}
u_{\eta}(\pmb{x})
  = \frac{1}{\sqrt{2\omega}}\, v_{k}^{(\mp)}(\pmb{x})
  = e^{-i\omega t}\, \phi_{\omega}^{(\mp)}(x),
\end{align}
where the spatial mode function $\phi_{\omega}^{(\mp)}(x)$ satisfies
\begin{align}\label{modeeq}
\left( -\frac{d^{2}}{dx^{2}} + m_{\mathrm{eff}}^{2}(x) \right)
\phi_{\omega}^{(\mp)}(x)
= \omega^{2}\, \phi_{\omega}^{(\mp)}(x)
\end{align}
with $m_{\mathrm{eff}}^{2}(x)$ given in \eqref{effMassSq1}.
In the right asymptotic region the quantized scalar field can be written as
\cite{Ho:2024kzr}
\begin{align}\label{phi(x)}
\phi(\pmb{x})
  \simeq \int_{0}^{\infty} \frac{dk}{\sqrt{2\pi}}\, \frac{1}{\sqrt{2\omega}}
    \sum_{i=\pm} \biggl[
        b_{k}^{(i)}\, v_{k}^{(i)}(\pmb{x})
        + \bigl(b_{k}^{(i)}\bigr)^{\dagger}\!
          \bigl(v_{k}^{(i)}(\pmb{x})\bigr)^{*}
    \biggr],
\end{align}
where the frequency is given by $\omega = \sqrt{k^{2}+m_{0}^{2}}$.
Using \eqref{phi(x)} in the state $|0\rangle_{\mathrm{R}}^{\epsilon}$, the
Wightman function becomes
\begin{align}\label{G+xx2}
G_{\epsilon}^{+}(\pmb{x},\pmb{x}')
  \simeq {}_{\mathrm{R}}^{\epsilon}\!\langle 0 \,|\,
  \phi(\pmb{x})\,\phi(\pmb{x}')
  \,|\, 0 \rangle_{\mathrm{R}}^{\epsilon}
  \simeq \int_{0}^{\infty} \frac{dk}{4\pi\,\omega}
      \sum_{i=\mp}
      v_{k}^{(i)}(\pmb{x})\,
      \bigl(v_{k}^{(i)}(\pmb{x}')\bigr)^{*},
\end{align}
where we used the commutation relation
\begin{align}
[b_{k}^{(i)}, (b_{k'}^{(j)})^{\dagger}] \simeq \delta^{ij} \delta(k - k').
\end{align}
We now evaluate the right-hand side of \eqref{G+xx2} using \eqref{uw} and
\eqref{modeeq}.
Under the condition \eqref{paracon}, the coefficient of the
$e^{-b x}$ term in the mode functions $\phi_{\omega}^{(\mp)}(x)$ vanishes, and
their asymptotic form near the right region simplifies to
\begin{align}\label{phiw2}
\phi_{\omega}^{(\mp)}(x)
  = \frac{1}{\sqrt{2\omega}}
    \left(
        1
        - \frac{(1-\alpha)\,(1+5\alpha^{2})\, m_{0}^{2}}
               {4 b\, (b \mp i k)}\, e^{-2 b x}
        + \mathcal{O}\!\bigl(e^{-3 b x}\bigr)
    \right) e^{\pm i k x}.
\end{align}

Some comments are in order.
For the supersymmetric background with $\alpha = 0$, an analytic solution to
\eqref{modeeq} is known~\cite{Ho:2022omx}.  
For $0 < \alpha < 1$, no analytic solution is available, and we instead obtain
an approximate solution in the vicinity of the right asymptotic region,

\begin{align}\label{nparacon}
\phi_{\omega}^{(\mp)}(x) &= \frac{1}{\sqrt{2 \omega}} \left[1 - \frac{2(1-\alpha)m_0^2 \bigl(1- \frac{\xi b^2}{m_0^2}\bigr) }{b(b \mp 2 i k)}\, e^{- b x} \right. 
\\
&\hskip 0.2cm\left. - \frac{(1-\alpha) m_0^2 \left( -3 + \alpha +  \frac{4\xi b^2}{m_0^2} (1+ \alpha) - \frac{4(1-\alpha)m_0^2}{b(b \mp 2 i k)} \Bigl(1- \frac{\xi b^2}{m_0^2}\Bigr)^{2}   \right)}{4 b (b \mp ik)}\, e^{- 2 b x}  + \mathcal{O}\big( e^{- 3 b x}\big) \right] \, e^{\pm ikx}. \nonumber
\end{align}
Without the condition \eqref{paracon}, the term of the order $e^{- bx}$ in \eqref{nparacon} survives.
Then, our approximation scheme breaks down.

Substituting the mode functions $v_{k}^{(i)}$ obtained from \eqref{uw} and 
\eqref{phiw2} into \eqref{G+xx2}, we obtain an unrenormalized expression for 
$G_{\epsilon}^{+}(\pmb{x},\pmb{x}')$, which will be renormalized using the
Hadamard procedure.
In $(1{+}1)$ dimensions, the short-distance behaviour of the Wightman function is 
universally fixed and takes the Hadamard form
\begin{align}\label{G+xxp}
    G^{+}(\pmb{x},\pmb{x}')
    = \frac{1}{4\pi} \Bigl(
        V(\pmb{x},\pmb{x}')\, 
        \ln\!\bigl(\mu^{2}\,\sigma(\pmb{x},\pmb{x}')\bigr)
        + W(\pmb{x},\pmb{x}';\mu)
      \Bigr),
\end{align}
where \(V(\pmb{x},\pmb{x}')\) and \(W(\pmb{x},\pmb{x}';\mu)\) are smooth
biscalars, finite in the coincidence limit \( \pmb{x}' \to \pmb{x} \), and can
be expanded in powers of the Synge function $\sigma (\pmb{x}, \pmb{x}')$,
\begin{align}\label{VWnxxp}
&V(\pmb{x}, \pmb{x}') = \sum_{n=0}^{+\infty} V_n(\pmb{x}, \pmb{x}') \sigma^n(\pmb{x}, \pmb{x}'), 
\nonumber\\
&W(\pmb{x}, \pmb{x}';\mu) = \sum_{n=0}^{+\infty} W_n(\pmb{x}, \pmb{x}';\mu) \sigma^n(\pmb{x}, \pmb{x}').
\end{align}  
The function \(V(\pmb{x},\pmb{x}')\) is fixed entirely by the background
geometry together with the field equation, whereas \(W(\pmb{x},\pmb{x}';\mu)\)
incorporates both geometric contributions and the dependence on the particular
quantum state.  
The constant \(\mu\), which has dimensions of mass, is introduced to render the 
argument of the logarithm dimensionless and is normalized so that the Minkowski
vacuum carries vanishing vacuum energy.  
The Hadamard coefficients $V_{n}(\pmb{x}, \pmb{x}')$ are obtained by solving
recursion relations along the geodesic connecting $\pmb{x}$ and $\pmb{x}'$.
By contrast, for the coefficients $W_n(\pmb{x}, \pmb{x}';\mu)$ the leading
term $W_0(\pmb{x}, \pmb{x}';\mu)$ is not fixed by these recursion relations and
encodes the choice of quantum state. For further discussion, see
\cite{Decanini:2005eg} or Appendix~A of \cite{Ho:2024kzr}. 
Here, $G_{\epsilon}^{+}(\pmb{x},\pmb{x}')$ in \eqref{G+xx2} is valid only around $x_\epsilon$ only,  whereas $G^{+}(\pmb{x},\pmb{x}')$ in \eqref{G+xxp} applies throughout the entire spacetime region.

The renormalized Wightman function around $x_\epsilon$ is then given by 
\begin{align}\label{renWM0}
G^{+}_{\text{ren}} (\pmb{x}, \pmb{x}'; \mu)
  =  \frac{1}{4 \pi}  W (\pmb{x}, \pmb{x}';\mu)
  = G_\epsilon^{+} (\pmb{x}, \pmb{x}')  
    - \frac{1}{4 \pi} V(\pmb{x}, \pmb{x}')
      \ln  \bigl[ \mu^2\sigma (\pmb{x}, \pmb{x}') \bigr].
\end{align}
Our aim is to compute the VEV of the energy--momentum tensor near the right
asymptotic limit using \( G^{+}_{\mathrm{ren}}(\pmb{x}, \pmb{x}'; \mu) \). 
Since \( G_{\epsilon}^{+}(\pmb{x}, \pmb{x}') \) in the right-hand side 
of~\eqref{renWM0} can already be obtained from \eqref{G+xx2} and 
\eqref{phiw2},  which is simply given in terms of the Minkowski Synge function $\sigma_{\rm M}$ by  $K_{0}(m\sqrt{2\sigma_{\rm M}})/2\pi$ in our approximation scheme (See~\cite{Ho:2024kzr}), 
the remaining task is to determine 
\( V(\pmb{x}, \pmb{x}') \) and \( \sigma(\pmb{x}, \pmb{x}') \) 
in the right asymptotic limit.

In the metric~\eqref{metricModified}, the spacelike Synge function is given by
\begin{align} \label{synge0}
2\sigma (\pmb{x},\pmb{x}') 
&=  -(\Delta t)^{2} + (\Delta x)^{2}  
 \nonumber\\
&\quad - 2 (1-\alpha)\, e^{-\frac{b}{2}(x+x')} 
 \Big[
   -(\Delta t)^{2} + (\Delta x)^{2}   
   - \frac{b^{2}}{24} (\Delta t)^{2} (\Delta x)^{2}  
   + \frac{b^{2}}{24}   (\Delta x)^{4} 
 \Big] 
 + \cdots ,
\end{align}
where $\Delta t \equiv t' - t$ and $\Delta x \equiv x' - x$, and the ellipsis
denotes higher-order terms involving combinations of $\Delta t$ and $\Delta x$. 
The procedure for computing the Synge function is described in detail in 
Appendix~\ref{appendixA}. 
In particular, we have obtained the general form of the Synge function with a
conformal factor $e^{2\omega} = e^{2 \omega (t,x)}$
for several leading terms in $(1+1)$ dimensions.
 
Without imposing the condition \eqref{paracon}, we now compute the biscalar $V(\pmb{x}, \pmb{x}')$ up to 
$\mathcal{O}(e^{-b x})$.  
The first two Hadamard coefficients~\cite{Decanini:2005eg} are
\begin{align} \label{hadamardcoeff}
V_0(\pmb{x}, \pmb{x}') 
&= -1 
- \frac{1}{12}\,
\mathcal{R}_{\alpha\beta}(\pmb{x})\,
\nabla_{\pmb{x}}^{\alpha} \sigma\,
\nabla_{\pmb{x}}^{\beta} \sigma
+ \mathcal{O}(\sigma^{3/2}), 
\nonumber \\
V_1(\pmb{x}, \pmb{x}')
&= -\frac{m_0^{2}}{2}
- \frac{\mathcal{R}(\pmb{x})}{4}
\left(2\xi - \frac{1}{3}\right)
+ \mathcal{O}(\sigma^{1/2}) .
\end{align}
Substituting the metric~\eqref{metricModified} and the Synge function
\eqref{synge0} into \eqref{hadamardcoeff}, we obtain
\begin{align}
V_0 &= -1 
- \frac{b^{2}}{12}(1-\alpha)\,
\big( -(\Delta t)^{2} + (\Delta x)^{2} \big)\,
e^{-b x}
+ \mathcal{O}(e^{-2 b x}),
\nonumber \\
V_1 &= -\frac{m_0^{2}}{2}
- \frac{b^{2}}{2}
\left(2\xi - \frac{1}{3}\right)
(1-\alpha)\,
e^{-b x}
+ \mathcal{O}(e^{-2 b x}),
\end{align}
where we used the Ricci tensor components in the metric
\eqref{metricModified},
\begin{align}\label{RACS}
\mathcal{R}_{tt} &= - b^{2}(1-\alpha)\, e^{-b x} 
  + \mathcal{O}(e^{-2 b x}), \nonumber \\
\mathcal{R}_{xx} &= \;\, b^{2}(1-\alpha)\, e^{-b x} 
  + \mathcal{O}(e^{-2 b x}), \\
\mathcal{R}_{tx} &= 0, \qquad
\mathcal{R} = 2b^{2}(1-\alpha)\, e^{-b x}
  + \mathcal{O}(e^{-2 b x}).
\nonumber
\end{align}
From these results, we obtain 
\begin{align}\label{V(xxp)}
V(\pmb{x}, \pmb{x}') 
&= -1 
- \frac{m_0^{2}}{4}\,
\big( -(\Delta t)^{2} + (\Delta x)^{2} \big)
 \\[1mm]
&\quad
+ \frac{1}{2}\big(m_0^{2} - \xi b^{2}\big)(1-\alpha)\,
\big( -(\Delta t)^{2} + (\Delta x)^{2} \big)\,
e^{-b x}
+ \mathcal{O}(e^{-2 b x}).  
\end{align}
Combining \eqref{G+xx2}, \eqref{phiw2}, \eqref{synge0}, and \eqref{V(xxp)}, we
finally obtain the renormalized Wightman function under the AWS condition
\eqref{paracon} up to  $\mathcal{O} (e^{- b x})$,
\begin{align}\label{G+ren}
G^{+}_{\rm ren} ({\pmb x}, {\pmb x}';\mu) 
&= \frac{1}{4\pi} \left(1 + \frac{m^2_{0}}{2} \sigma_{\rm M} \right) \ln \left( \frac{\mu^2}{2m^2_{0}} \right) 
 + \frac{1}{2\pi} \left(-\gamma + \ln 2\right) 
+ \frac{m^2_{0}}{4 \pi} \left(1 - \gamma + \ln 2\right) \sigma_{\rm M}  \nonumber \\[1mm]
&\quad - \frac{1-\alpha}{2\pi} \left(1 +  \frac{b^2}{24} (\Delta x)^2 + \frac{m^2_{0}}{2}\sigma_{\rm M} \right) 
e^{-b \frac{x + x'}{2}} + \cdots, 
\end{align}
where  $2\sigma_{\rm M} \equiv  -(\Delta t)^2 + (\Delta x)^2 $ and the ellipsis denotes higher-order terms in
$\Delta t$, $\Delta x$, and $e^{-b (x + x')/2}$.

\subsection{Quantum energy-momentum tensor and trace anomaly}
\label{subsec4.2}

We now compute the quantum energy--momentum tensor using the renormalized
Wightman function $G^{+}_{\mathrm{ren}}(\pmb{x}, \pmb{x}';\mu)$ obtained in 
the previous subsection~\ref{subsec4.1}. 
One of our aims is to determine,
through explicit calculations, whether the VEV of  $T^{\mu}{}_{\mu}$ with respect to the vacuum state
$\lvert 0 \rangle_{\mathrm{R}}^{\epsilon}$ defined at $x_\epsilon$ 
vanishes asymptotically, or instead acquires an anomalous contribution.

\subsubsection{VEV of the energy-momentum tensor}
\label{subsec4.3}

To obtain the VEV of $T_{\mu\nu}$, we apply the point-splitting method
and use the renormalized Hadamard function introduced in
\eqref{renWM0} and \eqref{G+ren}.  
For the vacuum $\lvert 0\rangle_{\rm R}^{\epsilon}$, we have
\begin{equation}\label{Tmnv}
\langle T_{\mu\nu}(\pmb{x};\mu)\rangle_{\rm R}^{\epsilon}
 = \lim_{\pmb{x}'\rightarrow\pmb{x}}
   \mathcal{T}_{\mu\nu'}\, G^{+}_{\mathrm{ren}}(\pmb{x},\pmb{x}';\mu),
\end{equation}
where the relevant bi-differential operator in $(1+1)$ dimensions is
\begin{align} \label{diffBiV}
\mathcal{T}_{\mu\nu'} =\;&
(1-2\xi)\,\partial_{\mu}\partial_{\nu'}
+ g_{\mu\nu'}\!\left[\left(2\xi-\tfrac12\right)
    g^{\alpha\beta'}\partial_{\alpha}\partial_{\beta'}
    - \tfrac12 m_{0}^{2}
    + 2\xi g^{\alpha\beta}\nabla_{\alpha}\nabla_{\beta}\right]
\nonumber\\[2mm]
& - 2\xi\,\delta^{\mu'}_{\mu}\,\partial_{\mu'}\partial_{\nu'}
  + \tfrac14\, g_{\mu\nu'}\, P_{x},
\end{align}
with $P_{x} = -\Box_{x} + m_{0}^{2} + \xi\mathcal{R}$.  
As discussed in~\cite{Moretti:2001qh,Moretti:2021pzz}, this last term in the
right-hand side of \eqref{diffBiV} is not included in the standard
treatment~\cite{Birrell:1982ix}, since it does not appear in the classical
tensor \eqref{classicalStress}. If the last term in~\eqref{diffBiV} is omitted, the VEV 
 of energy-momentum tensor is no longer covariantly conserved.  
For more details in our specific setup, see
\cite{Ho:2024kzr}.

Inserting \eqref{G+ren} into \eqref{Tmnv} and simplifying the expression using  the condition~\eqref{paracon}, we obtain
\begin{align}    \label{Ttt0}
\langle T_{tt}\rangle_{\rm R}^\epsilon  &=   \frac{m_{0}^{2}}{4\pi} \Big[\frac{1}{2}-\gamma + 
   \ln 2 +\frac{1}{2} \ln \Big(\frac{\mu^2}{2m_{0}^2}  \Big) \Big]  \Big[ 1 -2\big(1-\alpha \big)\, e^{-bx}\Big] 
\nonumber\\
&\quad    -\frac{b^{2}}{12\pi}\,  \big(1-\alpha \big) e^{-bx} +   {\cal O} \Big(e^{-2 b x}\Big) \,,  \\[1mm]
\label{Txx}
\langle T_{xx}\rangle_{\rm R}^\epsilon  &=   \frac{m_{0}^{2}}{4\pi} \Big[-\frac{1}{2}+\gamma - 
   \ln 2 -\frac{1}{2} \ln \Big(\frac{\mu^2}{2m_{0}^2}  \Big) \Big]    \Big[ 1 -2\big(1-\alpha \big)\, e^{-bx}\Big]  +   {\cal O} \Big(e^{-2 b x}\Big) \,,   \\[1mm]
\langle T_{tx} \rangle_{\rm R}^\epsilon  &= \langle T_{xt}  \rangle_{\rm R}^\epsilon
   ={\cal O} \Big(e^{-2 b x}\Big) \,. \label{Ttx}   
\end{align}
It is worth to note that all terms with the coefficient $\xi$ cancel among themselves in the above results.
Since the scalar curvature has the form given in \eqref{RACS}, the expressions
above can be interpreted as an expansion valid in the regime where the
curvature is small, namely in the limit $x \to \infty$. One can show that the 
VEV of the energy--momentum tensor given in \eqref{Ttt0}, \eqref{Txx}, and
\eqref{Ttx} is covariantly conserved up to ${\cal O}\!\big(e^{- b x}\big)$,
irrespective of the value of $\mu$. Note also that the above results reduce to those of our previous supersymmetric case~\cite{Ho:2024kzr} when $\alpha=0$ and $\xi=1/4$ ({\it i.e.} $m_{0}=b/2$).  

 Following the
procedure adopted in~\cite{Ho:2024kzr}, 
we require that the VEV of the energy--momentum tensor vanish in the Minkowski
vacuum $|0\rangle_{\rm M}$, i.e.
\begin{align} \label{Mvev}
{}_{\rm M}\langle 0|\, T_{\mu\nu}\, |0\rangle_{\rm M} = 0.
\end{align}
In the Minkowski vacuum, the VEV of the energy--momentum tensor is
\begin{align}
\langle T_{tt}\rangle_{\rm M}  &=  
\frac{m_0^2}{4\pi } \left[\frac{1}{2}-\gamma 
+ \ln 2 
+ \frac{1}{2}\ln\!\left(\frac{\mu^2}{2 m_0^2}\right) \right],
\nonumber\\[4pt]
\langle T_{xx}\rangle_{\rm M}  &=  
\frac{m_0^2}{4\pi } \left[-\frac{1}{2}+ \gamma 
- \ln 2 
- \frac{1}{2}\ln\!\left(\frac{\mu^2}{2 m_0^2}\right) \right],
\nonumber\\[4pt]
\langle T_{tx}\rangle_{\rm M} &= \langle T_{xt}\rangle_{\rm M} = 0.
\end{align}
Thus, in order to achieve the requirement \eqref{Mvev}, we must choose
\begin{align}\label{rensc}
\mu = \frac{m_0}{\sqrt{2}}\, e^{\gamma - \frac{1}{2}}.
\end{align}
Substituting this value of $\mu$ into \eqref{Ttt0}, \eqref{Txx}, and
\eqref{Ttx} yields the asymptotic form
\begin{align}
\langle T_{tt}\rangle_{\rm R}^{\epsilon}  
&=   -\frac{b^{2}}{12\pi}\,  (1-\alpha)\, e^{-b x}
   + {\cal O}\!\left(e^{-2 b x}\right), 
\nonumber\\[4pt] \label{Txx2}
\langle T_{xx}\rangle_{\rm R}^{\epsilon}  
&=   {\cal O}\!\left(e^{-2 b x}\right), \\[4pt]
\langle T_{tx}\rangle_{\rm R}^{\epsilon} 
&= \langle T_{xt}\rangle_{\rm R}^{\epsilon} 
   = {\cal O}\!\left(e^{-2 b x}\right). 
\nonumber
\end{align}
This result represents the quantum effects of the state 
$|0\rangle_{\mathrm{R}}^{\epsilon}$ defined at $x_\epsilon$, 
under the condition that the VEV of the energy--momentum tensor 
vanishes in the Minkowski vacuum. 
As shown in~\eqref{metricModified}, the background metric reduces to Minkowski
space when $\alpha = 1$, which is consistent with \eqref{Txx2}. 
According to our proposal~\cite{Kwon:2022fhv,Ho:2022omx, Ho:2024kzr}, we can identify 
\begin{equation}\label{idTmn}
\langle T_{\mu\nu}\rangle_{\rm IFT}^\epsilon
  = \langle T_{\mu\nu}\rangle_{\rm R}^\epsilon. 
\end{equation}

\subsubsection{Asymptotic Weyl anomaly}
\label{subsec4.4}

We now connect the formal AWS analysis of Section~\ref{sec3} with the explicit
results of the previous subsections, in order to exhibit the asymptotic trace
anomaly in the right asymptotic region. Using \eqref{Txx2} together with the curvature
scalar in \eqref{RACS}, we find
\begin{align}\label{eq:asymp-trace}
g^{\mu\nu}\langle T_{\mu\nu}(\pmb{x};\mu) \rangle_{\rm R}^{\epsilon}
 = \frac{1}{12\pi}\, b^{2}(1-\alpha)\, e^{-b x}
   + {\cal O}\!\left(e^{-2 b x}\right)
 = \frac{1}{24\pi}\,\mathcal{R}
   + {\cal O}\!\left(e^{-2 b x}\right),
\end{align}
which is precisely the asymptotic anomaly form anticipated in Section~\ref{sec3}.

Within the Hadamard renormalization scheme, a general two-dimensional scalar
field with the non-minimal coupling in \eqref{actionFTCS} satisfies the local relation \cite{Decanini:2005eg}
\begin{equation}
  g^{\mu\nu}\langle T_{\mu\nu}(x;\mu)\rangle
   = \bigl(\xi \Box - m_{0}^{2}\bigr)\,
      G^{+}_{\rm ren}(x,x;\mu)
     + \frac{1}{24\pi}\,\mathcal{R} \, .
\label{TraceHadamard}
\end{equation}
In our setup the renormalized Wightman function \eqref{G+ren} obeys
\begin{equation}
G^{+}_{\rm ren}(x,x;\mu)
 = -\frac{1}{2\pi}(1-\alpha)e^{-bx}
   + {\cal O}(e^{-2bx}) \,,
\end{equation}
so that the state-dependent term
$(\xi\Box-m_{0}^{2})G^{+}_{\rm ren}$ vanishes up to  ${\cal O}(e^{-2bx})$ in the right asymptotic region. 
Consequently, the purely
geometric curvature term in~\eqref{TraceHadamard} provides the leading
contribution to $\langle T^\mu_{~\mu}\rangle$.  This gives an explicit realization of the
asymptotic Weyl anomaly for the class of vacua considered here and confirms
the formal argument of Section~\ref{sec3}.

\subsection{Unruh-like effect in the bubble-wall expansion and IFT implication}
\label{subsec4.5}

Finally, we comment on the physical interpretation of the energy-momentum tensor and its
implications in the IFT description. The result that 
$\langle T_{tt} \rangle_{\mathrm{R}}^{\epsilon} < 0$ in \eqref{Txx2} suggests
that an observer located slightly away from the right asymptotic region
measures a negative local energy density with respect to the Minkowski
vacuum. By analogy with the Rindler case, this can be interpreted as indicating
that such an observer would detect a thermal-like distribution of particles
associated with the field, whereas the right Minkowski vacuum remains empty
for inertial observers at spatial infinity. 
Therefore, we interpreted this result as an Unruh-like effect. 
See \cite{Kwon:2022fhv,Ho:2022omx, Ho:2024kzr}.

In studies of first-order electroweak phase transitions, the growth of bubbles has long been analysed~\cite{Coleman:1977py, Callan:1977pt, Linde:1981zj, Moore:1995ua, Bodeker:2017cim, Azatov:2020ufh, Kubota:2024wgx, Ho:2024kzr, Kubota:2025avm}. Within thick-wall descriptions, models employing
spatially varying masses are standard, and we adopt the IFT with the mass profile given in \eqref{effMassSq1}.  
Here we focus solely on the vacuum contribution to the resistance acting on the
wall.   Writing the
total pressure as $P_{\rm tot}=-\Delta V+\Delta P$, only the latter can survive in
the wall rest frame, and in our zero-temperature model the classical contribution vanishes. Thus the relevant pressure is purely quantum. This bubble wall expansion can be modeled by our scalar IFT~\cite{Ho:2024kzr, Kubota:2024wgx, Kubota:2025avm}. Then, using the identification \eqref{idTmn}, 
near the Higgs-side region of the wall, we can see that the pressure difference becomes
\begin{equation}
\Delta P_\epsilon
  = \langle T_{xx}\rangle_{\rm IFT}^\epsilon -
    \langle T_{xx}\rangle_{\rm IFT}^{\rm H},
\end{equation}
with the Higgs-side contribution being zero in our prescription, where $\langle T_{xx}\rangle_{\rm IFT}^{\rm H} \equiv \langle T_{xx}\rangle_{{\rm M}}$.
From the asymptotic form of $\langle T_{xx}\rangle_{{\rm R}}^\epsilon$ in \eqref{Txx2},  we obtain\footnote{In our previous work~\cite{Ho:2024kzr}, due to a computational oversight, we reported a positive
pressure contribution at the leading order ($e^{-b x}$). This has been corrected
in the recent revision.
}
\begin{equation}
\Delta P_\epsilon
  = {\cal O}(e^{-2bx}) .
\end{equation}
Therefore, from the IFT perspective, although the quantum vacuum produces an
Unruh-like contribution as the leading effect (of order $e^{-b x}$), the
positive (or negative) pressure that would oppose (or enhance) the expansion of
the bubble wall vanishes at leading order. In other words, the effect of the
pressure difference $\Delta P$ begins to contribute only at least at the
next-to-leading order ($e^{-2b x}$). More detailed analysis of this issue will be deferred
to future work.

\section{Conclusions}
\label{sec5}

Modern understanding of  high energy physics is based on quantum field theory(QFT), which is regarded as an effective theory of the more fundamental theory,  string/M-theory. Nevertheless, a (effective) field theory has quite versatile applicability and makes  various conceptual and practical achievements. Therefore, it is  still meaningful to explore various extensions of QFT. In this regard, one of the crucial ingredients of relativistic QFT is the existence of a rather large global symmetry: Poincar\'e symmetry. However, in some low energy regimes,  there are situations lacking a large global symmetry. For instance, there would be  background fields which can be regarded as  non-dynamical. The well-known example of these situations is QFT with a fixed non-dynamical curved background spacetime.  In this case the algebraic approach has been adopted to overcome various conceptual issues,   reconciling the apparent coordinate dependence of the choice of Hilbert spaces. IFT explored in our recent works~\cite{Kwon:2022fhv,Ho:2022omx, Ho:2024kzr} is another example, which have a similar property with FTCS. But, the absence of general covariance renders IFT to be more difficult to understand its quantum aspects, even though it  would have some applications in various low energy regimes.

Because of the absence of a large global symmetry, there is no consensus how to quantize IFT in   generic situations. Though we have presented a certain way to accomplish the quantization procedures in a specific setup~\cite{Ho:2024kzr,Kwon:2022fhv,Ho:2022omx}, the complete answer to the quantization remains as an open issue. In this paper, we continue to explore the quantization of IFT and provide its interpretation. The main results in this paper are two folded. First, we  establish  the existence of a hidden approximate asymptotic Weyl-like symmetry, designated as AWS, in a specific FTCS, and provide its interpretation within the IFT framework. This observation explains our previous result on the anomaly expression of the trace of energy-momentum tensor VEV even in a certain non-conformal IFT, as shown in  \eqref{Txx2} and~\eqref{eq:asymp-trace}. A crucial element for this result is the choice of a physically appropriate vacuum. In Section~\ref{sec4}, we have provided how to choose approximately such a vacuum by using canonical formalism (See the paragraph below~\eqref{twopf}). From the algebraic viewpoint, this   corresponds to choosing the vacuum such that the first term in~\eqref{TraceHadamard} should vanish at the relevant order.  In this regard, we would like to emphasize again that  AWS is not exact but just asymptotic, which is valid only for a specific profile of the background metric.  

Second, we address concerns  in the  FTCS framework, regarding the validity of Hadamard renormalization method in a spacetime with a null singularity.  Since our one-parameter extension of    the metric  for  $\alpha \neq 0$ admits  no singularity, we   eliminated   some doubts about the applicability of the Hadamard renormalization. This in turn shows us that  the physical interpretation of our results in IFT is more reliable. 

Now, we would like to provide some comments about IFT under consideration with respect to another inhomogeneous theory with a different perspective. In  a recent work~\cite{Fan:2019upv,Lapierre:2019rwj,Lapierre:2024lga,Erdmenger:2025chu}, the so-called driven inhomogeneous CFT~\cite{deBoer:2023lrd,Oblak:2016eij} is explored in conjunction with a curved spacetime metric and the AdS/CFT correspondence. In this work, the inhomogeneity is introduced into the global Hamiltonian while local conformal operator structure is preserved. In other words, a Hamiltonian is deformed by a inhomogeneous function while a tensor operator or a corresponding state  can still be classified by a conformal weight. It is interesting to note that the interpretation of the driven inhomogeneous CFT  in terms of CFT on a certain curved background is quite similar to our case. Furthermore, the choice of  a specific renormalization scheme  is emphasized in~\cite{Erdmenger:2025chu}. This  is also similar with our case in the sense that a specific renormalization scheme is required to convert FTCS to IFT.  However, there is one crucial difference. In the case of inhomogeneous CFT, there is a large conformal symmetry allowing the famous state-operator correspondence. Therefore,  there seems  no issue on the choice of Hilbert space in the conversion from the driven inhomogeneous CFT to CFT on curved spacetime. In some sense, the large symmetry allows a preferred choice of vacuum in Hilbert space. This is not the case in our examples and one of the main issues in our approach.  In our previous paper~\cite{Kwon:2022fhv, Ho:2022omx, Ho:2024kzr}, we have provided one way to choose an observer-dependent (global) vacuum in the context of the algebraic approach, which is also adopted in this paper. All these results tell us that a specific renormalization scheme with a careful choice of a certain vacuum is quite crucial to understand the quantization of IFT.

There are various directions to pursue in the future. One of urgent directions is to understand the more concrete nature of AWS. It would be very interesting whether it exists in other field theory.  Since our observation about this new type symmetry is a bit non-conventional in the sense that it is just asymptotically valid, it requires more clarification. It would also be interesting to analyze the left asymptotic region, where reflection modes
play an essential role.
Another issue   is related to fermionic IFT, which would have some applications in  condensed matter physics.  Especially, the fermionic zero modes seems to exist when it is realized as a superpartner of scalar IFT, which is also needed to be explored.    The finite temperature physics is still needed to be explored in the context of IFT, which may be related to non-equilibrium physics and the driven inhomogeneous CFT. 
\vskip 1cm

\section*{Acknowledgments}
This work was supported  in part by the National Research Foundation of Korea(NRF) grant with grant number RS-2019-NR040081 (O.K., J.H., and S.-A.P.), RS-2023-00208047 (J.H.), RS-2023-00208011 (S.-H.Y.),  and  in part by NRF
grant funded by the Korea government(MSIT) RS-2025-00553127 (O.K., J.H., S.-A.P.,  and S.-H.Y.).

\vskip 1cm

\begin{center} {\Large \bf Appendix}
\end{center}

\begin{appendix}

\section{Synge Function}\label{appendixA}

In this appendix, we provide some details about properties of Synge function relevant in the main text. 
To compute the renormalized (positive-frequency)  Wightman function and the VEV of energy-momentum tensor, we need the short-distance expansion of Synge function in terms of coordinates. 
Following Ref.s~\cite{Poisson:2011nh,Siemssen:2015owa}, we define Synge function as  a bi-tensor  along the geodesic  connecting two points $\pmb{x}'$ and $\pmb{x}$ 
\begin{equation} \label{}
\sigma( \pmb{x}',\pmb{x})  \equiv \frac{1}{2} (\lambda_{1}-\lambda_{0}) \int^{\lambda_{1}}_{\lambda_{0}} d\lambda  ~ g_{\mu\nu}t^{\mu}t^{\nu},,  \qquad t^{\mu} \equiv \frac{dx^{\mu}}{d\lambda}\,,
\end{equation}
where $\lambda_{0}$ denotes the affine parameter for the 'base point'  $\pmb{x}$ and $\lambda_{1}$ does for the field point $\pmb{x}'$. 
It is straightforward to see that this function is one half of the signed squared geodesic distance for two points  $\pmb{x}'$ and $\pmb{x}$. One of the essential property of Synge function is  
\begin{equation} \label{}
\sigma_{\mu}\sigma^{\mu} = \pm 2 \sigma\,, \qquad \sigma_{\mu} \equiv \nabla^{\mu}\sigma\,.
\end{equation}

In the following, we  would like to derive the following expansion of Synge function:
\begin{equation} \label{}
\sigma(\pmb{x}',\pmb{x}) = \sum_{m=0}^{\infty}\frac{1}{m!}\zeta_{\mu_{1}\mu_{2}\cdots \mu_{m}}(\pmb{x})\, \Delta x^{\mu_{1}}\Delta x^{\mu_{2}}\cdots \Delta x^{\mu_{m}}\,, 
\end{equation}
where $\Delta x^{\mu} \equiv (\pmb{x}'-\pmb{x})^{\mu}$ and $\zeta_{\mu_{1}\mu_{2}\cdots \mu_{m}} =\zeta_{(\mu_{1}\mu_{2}\cdots \mu_{m})} $   is defined recursively by\footnote{The parenthesis for indices denotes totally symmetrization of indices.}
\begin{align}    \label{SyngeRec}
  2(1-\ell) \zeta_{\mu_{1}\cdots \mu_{\ell}} &= -2 \ell\, \zeta_{\mu_{1}\cdots \mu_{\ell-1},\,  \mu_{\ell}} \nonumber \\
& \quad  + \sum_{k=2}^{\ell-2} {\ell \choose k}   ~ g^{\alpha\beta} \Big( \zeta_{(\mu_{1}\cdots \mu_{k},\,|\, \alpha} - \zeta_{\mu_{1} \cdots \mu_{k}\,|\,  \alpha} \Big)  \Big( \zeta_{\mu_{k+1}\cdots \mu_{\ell},\,|\,  \beta} - \zeta_{\mu_{k+1} \cdots \mu_{\ell} \,|\, \beta} \Big)  \,.  
\end{align}
with the initial data
\begin{equation} \label{}
\zeta =0\,, \qquad \zeta_{\mu} =0\,, \qquad \zeta_{\mu\nu} = g_{\mu\nu}\,.
\end{equation}

\noindent{\bf Derivation of the recursion relation}\\
Using $\sigma_{\alpha} \sigma^{\alpha} = \pm 2 \sigma$, we go ahead as follows. First, note that
\begin{equation} \label{}
\sigma_{\alpha} =  \sum_{n=1}^{\infty}\frac{1}{n!} \Big( \zeta_{\mu_{1}\mu_{2}\cdots \mu_{n},\, \alpha} - \zeta_{\mu_{1}\mu_{2} \cdots \mu_{n} \alpha} \Big) \Delta x^{\mu_{1}}\cdots \Delta x^{\mu_{n}}\,,
\end{equation}
where one may note that $\zeta_{\mu,\, \alpha}=0$. Here, ${ }'$ denotes the covariant derivative as $\zeta_{\mu,\, \alpha}\equiv \nabla_{\alpha}\zeta_{\mu}$.  Therefore, the first term in the right hand side is given by $-\zeta_{\mu \alpha}\Delta x^{\mu}$. Squaring this expression, we have 
\begin{align}    \label{}
\sigma_{\alpha} \sigma^{\alpha} &=  \sum_{m,n=1}^{\infty}\frac{1}{m!\, n!} ~ g^{\alpha\beta} \Big( \zeta_{\mu_{1}\mu_{2}\cdots \mu_{m},\, \alpha} - \zeta_{\mu_{1}\mu_{2} \cdots \mu_{m} \alpha} \Big)  \Big( \zeta_{\mu_{1}\mu_{2}\cdots \mu_{n},\, \beta} - \zeta_{\mu_{1}\mu_{2} \cdots \mu_{n} \beta} \Big)  \nonumber \\
& \qquad  \qquad  \qquad \qquad  \qquad \Delta x^{\mu_{1}}\cdots \Delta x^{\mu_{m}}\Delta x^{\mu_{1}}\cdots \Delta x^{\mu_{n}}\,,   \nonumber \\
&= 2\sum_{\ell=1}^{\infty}\frac{1}{\ell !}~ g^{\alpha\beta}(-\zeta_{\mu\alpha})\Big(\zeta_{\nu_{1}\nu_{2}\cdots \nu_{\ell},\, \beta} - \zeta_{\nu_{1}\nu_{2}\cdots \nu_{\ell} \beta}\Big) \Delta x^{\mu}\, \Delta x^{\nu_{1}}\cdots \Delta x^{\nu_{\ell}} \nonumber \\
& \quad+ \sum_{m,n=2}^{\infty}\frac{1}{m!\, n!} ~ g^{\alpha\beta} \Big( \zeta_{\mu_{1}\mu_{2}\cdots \mu_{m},\, \alpha} - \zeta_{\mu_{1}\mu_{2} \cdots \mu_{m} \alpha} \Big)  \Big( \zeta_{\mu_{1}\mu_{2}\cdots \mu_{n},\, \beta'} - \zeta_{\mu_{1}\mu_{2} \cdots \mu_{n} \beta} \Big)    \nonumber \\
& \qquad  \qquad  \qquad  \qquad \qquad \Delta x^{\mu_{1}}\cdots \Delta x^{\mu_{m}}\Delta x^{\mu_{1}}\cdots \Delta x^{\mu_{n}}\,,  \nonumber 
\end{align}
Note that the first and second lines in the right hand side of the above last equality can be reshuffled as 
\begin{align}
1st &= -2\sum_{\ell=1}^{\infty}\frac{1}{\ell !} \Big(\zeta_{\mu_{1}\mu_{2}\cdots \mu_{\ell},\, \mu_{\ell+1}} - \zeta_{\mu_{1}\mu_{2}\cdots \mu_{\ell} \mu_{\ell+1}}\Big) \Delta x^{\mu_{1}}\cdots \Delta x^{\mu_{\ell}}\,\Delta x^{\mu_{\ell+1}} \nonumber \\
&=-2\sum_{\ell=1}^{\infty}\frac{1}{\ell !} \Big(\zeta_{(\mu_{1}\mu_{2}\cdots \mu_{\ell},\, \mu_{\ell+1})} - \zeta_{\mu_{1}\mu_{2}\cdots \mu_{\ell} \mu_{\ell+1}}\Big) \Delta x^{\mu_{1}}\cdots \Delta x^{\mu_{\ell}}\,\Delta x^{\mu_{\ell+1}}\,,  \label{1st}\\
2nd &=   \sum_{m,n=2}^{\infty}\frac{1}{m!\, n!} ~ g^{\alpha\beta} \Big( \zeta_{\mu_{1}\mu_{2}\cdots \mu_{m},\, \alpha} - \zeta_{\mu_{1}\mu_{2} \cdots \mu_{m} \alpha} \Big)  \Big( \zeta_{\mu_{1}\mu_{2}\cdots \mu_{n},\, \beta} - \zeta_{\mu_{1}\mu_{2} \cdots \mu_{n} \beta} \Big)    \nonumber \\
& \qquad  \qquad  \qquad  \qquad \qquad \Delta x^{\mu_{1}}\cdots \Delta x^{\mu_{m}}\Delta x^{\mu_{1}}\cdots \Delta x^{\mu_{n}}  \nonumber \\ 
&= \sum_{\ell =4}^{\infty}  \sum_{k=2}^{\ell-2}\frac{1}{(\ell-k)!k!} ~ g^{\alpha\beta} \Big( \zeta_{(\mu_{1}\cdots \mu_{k},\,|\, \alpha} - \zeta_{\mu_{1} \cdots \mu_{k}\,|\,  \alpha} \Big)  \Big( \zeta_{\mu_{k+1}\cdots \mu_{\ell},\,|\,  \beta'} - \zeta_{\mu_{1}\mu_{2} \cdots \mu_{n} \,|\, \beta} \Big)   \nonumber \\
&  \qquad  \qquad  \qquad  \qquad \qquad  \qquad  \Delta x^{\mu_{1}}\cdots \Delta x^{\mu_{\ell}}     \label{2nd} 
\end{align}
Using the relation $\sigma_{\alpha}\sigma^{\alpha} = 2\sigma$, we should identify
\begin{equation} \label{}
1st + 2nd = 2\sigma = 2\sum_{\ell=2}^{\infty} \frac{1}{\ell !} ~\zeta_{\mu_{1}\cdots \mu_{\ell}}~  \Delta x^{\mu_{1}}\cdots \Delta x^{\mu_{\ell}}  \,.
\end{equation}
Using the reshuffled expressions in~\eqref{1st} and~\eqref{2nd} in the above equation and collecting the same order in $ \Delta x^{\mu_{1}}\cdots \Delta x^{\mu_{\ell}}$, we finally obtain (for $\ell >2$) the recursion relation~\eqref{SyngeRec}.
%
%
%

Now, it is straightforward to derive 
\begin{equation} \label{}
\zeta_{\mu\nu\rho} = 3g_{\alpha(\rho} \Gamma^{\alpha}_{\mu\nu)}\,, 
\end{equation}
and 
\begin{equation} \label{}
\zeta_{\mu\nu\alpha\beta} = \frac{4}{3}\partial_{(\beta}\zeta_{\mu\nu\alpha)} - \frac{1}{6} g^{\rho\sigma}\Big(\partial _{\rho}  \zeta_{(\mu\nu} - \zeta_{\rho(  \mu\nu} \Big) \Big( \partial _{|\sigma|}  \zeta_{\alpha\beta)} - \zeta_{|\sigma| \alpha\beta)} \Big) \,.
\end{equation}
In our two-dimensional case, it is sufficient to keep  the terms up to the order of $m=4$ in computing the Hadamard renormalization.

For the one-parameter extended metric in the main text
\begin{equation} \label{}
ds^{2} = \bigg(\frac{1+\alpha e^{-bx}}{1+e^{-bx}}\bigg)^{2} = 1-2 (1-\alpha) e^{-bx} + {\cal O}(e^{-bx})\,,
\end{equation}
we have 
\begin{align}    \label{}
\zeta_{ttx} & = \zeta_{txt} = \zeta_{xtt} = -b(1-\alpha) \, e^{-bx} + {\cal O}(e^{-2bx}) \,,    \nonumber \\
 \zeta_{xxx} & = \frac{be^{-bx}}{(1+e^{-bx})^{3}} = 3b(1-\alpha) \,e^{-bx} + {\cal O}(e^{-2bx}) \,,  \nonumber 
\end{align}
and
\begin{align}    \label{}
\zeta_{ttxx} &= \zeta_{txtx}=\zeta_{txxt}=\zeta_{xttx}=\zeta_{xtxt} =\zeta_{xxtt}= \frac{2b^{2}}{3}(1-\alpha)\, e^{-bx} + {\cal O}(e^{-2bx}) \,,   \nonumber \\
\zeta_{xxxx} &=   -4 b^{2}(1-\alpha)\, e^{-bx}  + {\cal O}(e^{-2bx}) \,. \nonumber 
\end{align}
Then, finally we have 
\begin{align}    \label{}
2\sigma (\pmb{x}',\pmb{x}) 
&=  -(\Delta t)^{2} + (\Delta x)^{2}  - 2 (1-\alpha)\, e^{-\frac{b}{2}(x+x')} \Big[ -(\Delta t)^{2} + (\Delta x)^{2}  \nonumber \\
& \hskip6.5cm - \frac{b^{2}}{24} (\Delta t)^{2} (\Delta x)^{2}  + \frac{b^{2}}{24}   (\Delta x)^{4}   \Big]   + \cdots\,,\nonumber 
\end{align}
which is used in the main text. Note that one can obtain the Synge function expansion of the supersymmetric background metric in~\cite{Ho:2024kzr} by taking $\alpha=0$.

Just for a reference, we present the result of Synge function expansion for a generic conformal  case in (1+1) dimensions  
\begin{align}
	ds^{2}=e^{2\omega(t,x)}(-dt^{2}+dx^{2})\,.
\end{align}
In this case, we have 
\begin{align}
	2\sigma(\pmb{x}',\pmb{x})
	&= \,(-\Delta t^{2} + \Delta x^{2}) e^{2\omega}
\Bigg[
1 + \dot{\omega}\,\Delta t + \omega'\,\Delta x
+ \frac{1}{3}\bigl(\dot{\omega}' + 2\dot{\omega}\,\omega'\bigr)\Delta t\,\Delta x \nonumber \\
&\qquad\qquad\qquad\qquad\qquad + \frac{1}{6}\bigl(\ddot{\omega} + 2\omega'{}^{2}\bigr)\Delta t^{2}
+ \frac{1}{6}\bigl(\omega'' + 2\dot{\omega}^{2}\bigr)\Delta x^{2}
+ \cdots
\Bigg]\,,
\end{align}
where $\dot{}\equiv\frac{\partial}{\partial t}$.

\section{Non-supersymmetric Backgrounds in FTCS}
\label{appendixB}

In this work, we consider the static metric given in \eqref{metricModified}. 
For $\alpha = 0$, this metric reduces to the supersymmetric background on the FTCS 
that we analyzed in our previous paper~\cite{Ho:2022omx}, whereas for $\alpha = 1$ it becomes the flat metric. 
Therefore, the parameter $\alpha$ provides a continuous interpolation between the supersymmetric background 
and the flat background. Since the flat background can also be regarded as a supersymmetric background, 
the static metric \eqref{metricModified} preserves supersymmetry at $\alpha = 0$ with 
$\mathcal{N} = \big( \tfrac{1}{2}, \tfrac{1}{2} \big)$ or at $\alpha = 1$ with $\mathcal{N} = (1,1)$. However, the FTCS \eqref{actionFTCS} we are interested in generically breaks supersymmetry completely 
on the static background \eqref{metricModified} for $0 < \alpha < 1$. The reason is as follows.

A systematic framework for constructing supersymmetric field theories 
on curved backgrounds was developed in \cite{Festuccia:2011ws}. 
In the rigid limit $M_{\rm Pl} \to \infty$, the gravity multiplet becomes 
non-dynamical, whereas the matter multiplets remain dynamical and 
continue to exhibit global supersymmetry provided certain consistency 
conditions are satisfied. Within this setup, the requirement that the 
supersymmetry variation of the gravitino vanish leads to the generalized 
Killing spinor equation
\begin{equation}\label{gKSP}
  \nabla_\mu \epsilon(\pmb{x})
  = M_\mu(\pmb{x})\, \epsilon(\pmb{x}) \,,
\end{equation}
where $\epsilon$ is the supersymmetry parameter and 
$M_\mu(\pmb{x})$ is a background-dependent matrix acting on spinor 
indices, determined entirely by the bosonic components of the 
non-dynamical gravity multiplet.
See also \cite{Kwon:2021flc} for  $\mathcal{N} = \big( \frac{1}{2}, \frac{1}{2}\big)$ for general space-dependent parameters in (1+1)-dimensional Minkowski space. 

In our previous work~\cite{Ho:2022omx}, we considered a special class of 
backgrounds admitting supersymmetric FTCS solutions, characterized by  
\( M_\mu = \tfrac{1}{2} f(\mathcal{R}) \gamma_\mu \).  
For the particular choice \( f(\mathcal{R}) = (\xi/m_0)\,\mathcal{R} \), the 
bosonic sector of the FTCS reduces to the action in~\eqref{actionFTCS}.  
In this setup, for the conformally flat metric~\eqref{metric}, the warp factor 
\(\omega(x)\) satisfies the differential equation
\begin{align}\label{warpeqn}
  \omega'' + \frac{m_0}{2\xi} e^{\omega}\, \omega' = 0\,,
\end{align}
and provides a solution to the generalized Killing spinor equation~\eqref{gKSP}, 
preserving \(\mathcal{N} = \bigl(\tfrac{1}{2}, \tfrac{1}{2}\bigr)\) supersymmetry.
It is straightforward to verify that the modified metric~\eqref{metricModified} 
fails to satisfy the condition~\eqref{warpeqn} for generic values of \(\alpha\), 
except in the case \(\alpha = 0\).  
Consequently, the FTCS action~\eqref{actionFTCS} cannot be supersymmetrized on 
the background geometry~\eqref{metricModified} for \(0 < \alpha < 1\).

\end{appendix}

\end{document}